\documentclass[a4paper]{jpconf}
\usepackage{mkfig}
\usepackage{graphicx}
\begin{document}
\title{An Improved Parametrized Representation of the Secondary
  He Neutral Flow in its Source Region}

\author{Brian E. Wood$^1$}

\address{$^1$ Naval Research Laboratory, Space Science Division,
  Washington, DC 20375, USA}

\ead{brian.wood@nrl.navy.mil}


\begin{abstract}

     Analysis of data from the {\em Interstellar Boundary EXplorer} (IBEX)
has revealed the presence of a flow of neutral helium through the inner
solar system that most likely emanates from the outer heliosheath, where
a distinct population of neutral He is produced by charge exchange
processes.  This secondary He flow has been modeled using codes designed
to study interstellar flows through the heliosphere, but a laminar flow
is not a good approximation for the outer heliosheath.  I present a
simple parametrization for a more appropriate divergent flow, and
demonstrate how the secondary He particles might provide a means to
remotely measure the divergence of the ISM flow around the heliopause.

\end{abstract}

\section{Introduction}

     Launched in 2008, the {\em Interstellar Boundary EXplorer} (IBEX)
mission is designed to study various populations of neutral particles
flowing through the inner solar system (McComas et al.\ 2009).  One
such population is that of the local interstellar medium, which is
partly neutral.  The plasma component of the ISM will be deflected
around the heliopause, but the neutral component can penetrate into the
inner heliosphere, where these particles can be observed and provide
useful diagnostics on the characteristics of the
undisturbed ISM just outside the heliosphere.  Of particular interest
are neutral He atoms.  Unlike neutral H, for example, neutral He has
a very low charge exchange cross section and will therefore reach
the inner solar system with its flow relatively unaffected by
particle interactions.  Observations of neutral He have therefore
been crucial for measuring the local ISM flow vector, based on both
IBEX data and similar observations from the older {\em Ulysses}
mission (Witte 2004; Bzowski et al.\ 2012, 2014; McComas et al.\ 2015;
Wood et al.\ 2015).

     However, in the process of studying IBEX observations of the
ISM neutral He flow, a second component of neutral He was discovered,
which was dubbed the ``Warm Breeze'' component (Kubiak et al.\ 2014).
A signature of this secondary component may also be present in
{\em Ulysses} data (Wood et al., in preparation).
Based on continuing analysis of these data, it is becoming
increasingly certain that the second component is due
to a population of neutral He created by charge exchange outside the
heliopause (i.e., the outer heliosheath) (Kubiak et al.\ 2016).
Whether a bow shock exists or not, the ISM
plasma outside the heliopause will be heated, compressed, and
deflected (e.g., Zank et al.\ 2013).  The secondary He component
observed by IBEX is presumed to be created by charge exchange with
this heated, compressed, and deflected ISM plasma flow, predominantly by
the ${\rm He^{0}} + {\rm He}^{+}\rightarrow {\rm He}^{+} + {\rm He^{0}}$
reaction, which is important due to the significant
abundance of both ${\rm He^{0}}$ and ${\rm He}^{+}$ in the ISM
(Bzowski et al.\ 2012; M\"{u}ller et al.\ 2013).

     Initial analyses of the secondary He component have typically
used the same techniques used to analyze the primary ISM component,
assuming a Maxwellian laminar flow from infinity, affected primarily
by the Sun's gravity and photoionization as the neutrals approach
the inner solar system.  The most recent such analysis found a
secondary flow of $V_{\infty}=11.3$ km~s$^{-1}$ towards ecliptic
coordinates ($\lambda$,$\beta$)=($71.6^{\circ}$,$-12.0^{\circ}$), with
a temperature of $T=9500$~K, and an abundance of $5.7\%$ of the
primary ISM component (Kubiak et al.\ 2016).

     However, the reliability of these measurements are questionable,
because although the ``laminar flow from infinity'' assumption is
appropriate for the primary component, it is a poor approximation for
the outer heliosheath source region of the secondary component, where
the flow could more accurately be described as a ``divergent flow from
about 150~AU.''  The inferred temperature of the secondary neutrals is
particularly suspect, as $T=9500$~K is much lower than expected for
the outer heliosheath, where $T\approx 20,000$~K is more likely (e.g.,
Izmodenov \& Alexashov 2015).  The goal of this article is to present
an alternative flow parametrization that is more appropriate for the
secondary component, and to show how the divergence of that flow
affects the observed He atoms at 1~AU.

\section{Parametrizing a Divergent Flow}

\begin{figure}[t]
\plotfiddle{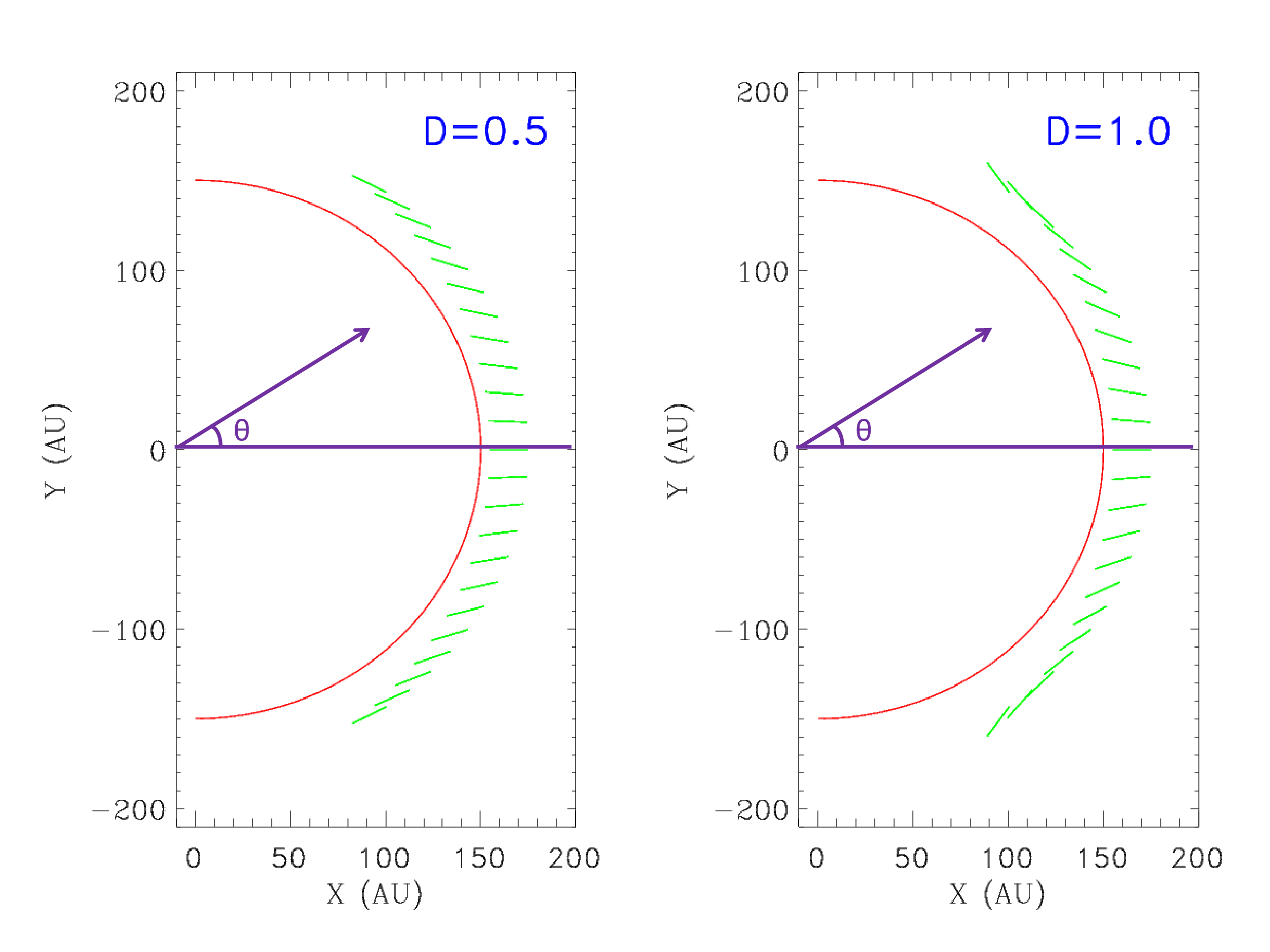}{3.2in}{0}{55}{55}{-200}{-25}
\caption{Representations of divergent flows from a distance of 150~AU
  (red line) for divergence parameters D=0.5 and D=1.0.}
\end{figure}
     We define our flow in a coordinate system with the Sun at the
origin and the x-axis pointing in the upwind direction of the ISM
flow (i.e., the stagnation axis), as in Figure~1.  For our flow from
the outer heliosheath, we define an outer boundary at a distance of
150~AU, as this is roughly the distance to the heliopause.  The most
important difference between the outer heliosheath flow and that
of the ISM is the divergence of the flow around the heliopause.
If $\psi$ is the angle of the flow from the x-axis, then we define
a divergence parameter $D$ such that $\psi=D\times \theta$, with $\theta$
being the angle defined in Figure~1.  In practice, we assume a
maximum value of $\psi=80^{\circ}$ regardless of $D$.  Figure~1 shows
the flow patterns associated with $D=0.5$ and $D=1.0$.

     In order to illustrate how $D$ affects the shape of a He beam
observed at 1~AU, we use a simple flow model in two-dimensions,
where the flow from the outer boundary is assumed to be affected
only by solar gravity, and the flow is modeled using a
routine that computes particle trajectories using simple numerical
integration (e.g., Wood et al.\ 2002).    We assume the velocity
distribution function (VDF) at the outer boundary (at $r=150$~AU) is
Maxwellian.  We then compute the resulting particle distribution
observed at (X,Y)=(1,0) in the coordinate system shown in Figure~1,
where for simplicity we are assuming the observer is on the
stagnation axis at 1~AU.

\begin{figure}[t]
\plotfiddle{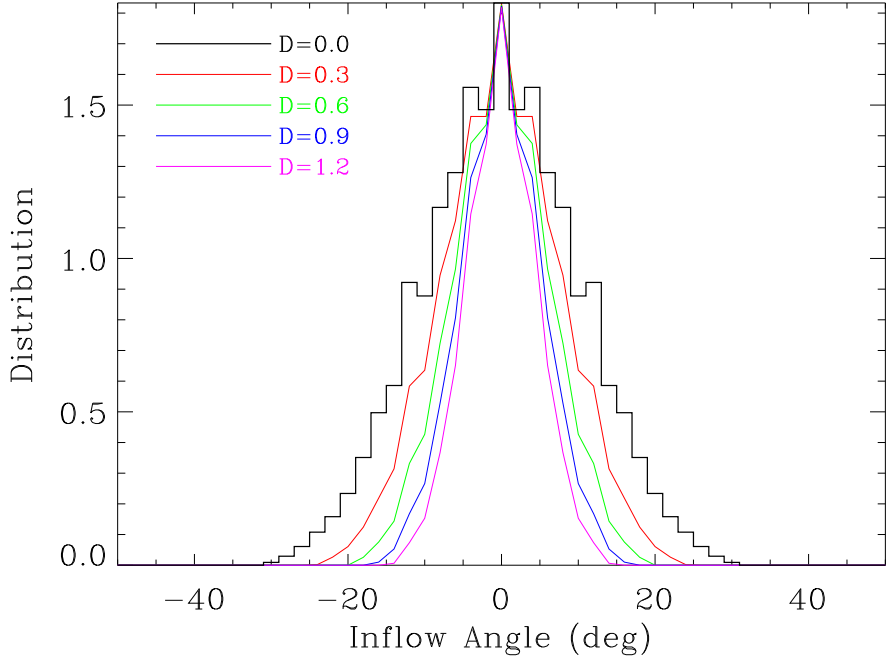}{3.2in}{90}{55}{55}{200}{-45}
\caption{Models of the He distribution observed at (X,Y)=(1,0), assuming
  a divergent flow from 150~AU with $V=20$ km~s$^{-1}$, $T=10^4$~K, and
  five values for the divergence parameter, in the range $D=0.0-1.2$.}
\end{figure}
     The black line in Figure~2 shows the observed distribution
for a flow with $V=20$ km~s$^{-1}$ and $T=10^4$~K, assuming
no divergence (e.g., $D=0$).  Limitations in the sampling
resolution of the VDF in the simulation lead to some noise in
the observed He distribution.  Colored lines in Figure~2 show
how the observed distribution narrows as $D$ is increased from
0 to 1.2.  Beam widths, $W$, can be quantified as the
full-width-at-half-maxima of Gaussians fitted to the distributions.
In Figure~2, increasing $D$ from 0 to 1.2 leads to a decrease in
width from $W=24.3^{\circ}$ to $W=11.4^{\circ}$.

     The decrease of $W$ with increasing $D$ happens because as
the observer looks farther from the stagnation axis (i.e, as $\theta$
increases), the flow at the boundary becomes directed further away
from the line of sight relative to a laminar flow, the result being
a faster decrease in observed flux with $\theta$ compared to a laminar
flow, and therefore a narrower He beam.  Decreasing $T$ would also
naturally narrow the observed distributions, as the VDF at the
source is narrower.  Thus, if $D$ is increased, one can preserve the
same $W$ by increasing $T$, illlustrating how assuming a divergent
flow would increase the $T=9500$~K measurement of Kubiak et al.\ (2016)
towards a more reasonable value.  The width will also be affected by
the velocity $V$, because as $V$ is decreased there is more dispersion
introduced into the observed distribution by the Sun's gravitational
effects.

\section{Remotely Measuring Flow Divergence}

     In order to further explore the dependence of $W$ on $D$, $T$,
and $V$; a number of trials like those in Figure~2 are computed,
assuming a range of $D$, $T$, and $V$ values, and measuring the $W$
values associated with them.  The observed ($D$,$T$,$V$,$W$) points
are fitted with the following power law relation:
\begin{equation}
W=C\left( \frac{V}{20} \right)^{\alpha} \left( \frac{T}{10^4} \right)^{\beta}
  \left( D+1 \right)^{\gamma},
\end{equation}
with $V$ in km~s$^{-1}$ and $T$ in K.  A least squares analysis is
used to determine the values of $C$, $\alpha$, $\beta$, and $\gamma$
that lead to the best fit to the measurements.  The best fit is shown
in Figure~3, with $C=24.1^{\circ}$, $\alpha=-0.84$, $\beta=0.52$, and
$\gamma=-0.91$.  The quality of the fit demonstrates that the
power law relation provides a reasonable approximation for how $W$
depends on $D$, $T$, and $V$.

\begin{figure}[t]
\plotfiddle{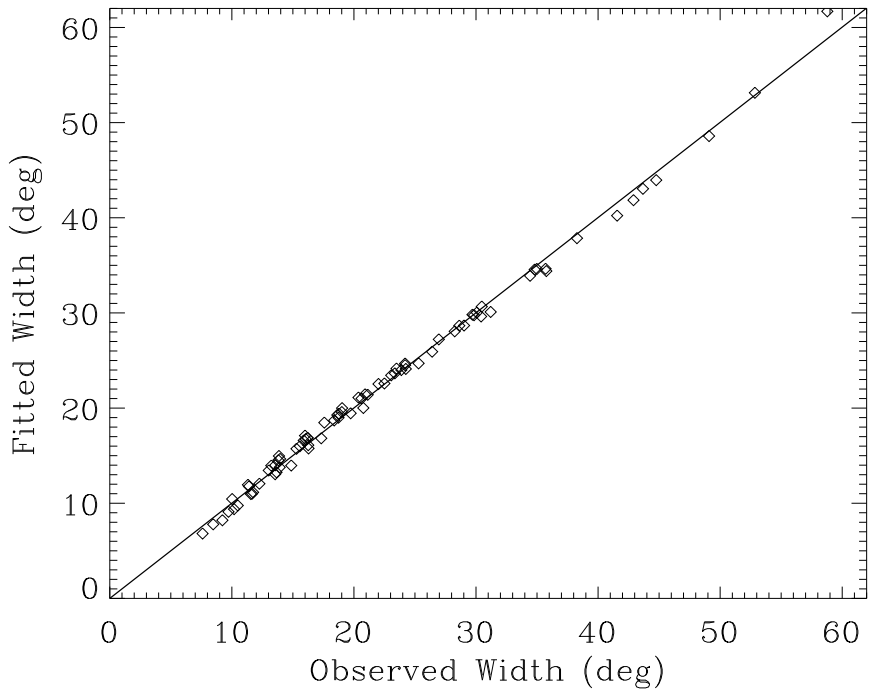}{3.2in}{90}{55}{55}{200}{-45}
\caption{Comparison between the He distribution widths predicted by the
  power law relation between $W$, $V$, $T$, and $D$ in equation~(1);
  and the observed widths based on numerical models, like those in
  Figure~2.}
\end{figure}
     Based on equation~(1), the Kubiak et al.\ (2016) fit to the
secondaries, ($D$,$T$,$V$)=(0,9500,11.3), suggests a He beam width of
$W=37.9^{\circ}$ in the context of our simple 2-D model geometry.  Global
heliospheric models suggest $V\approx 9$ km~s$^{-1}$ and
$T\approx 21,000$~K for the outer heliosheath (Kubiak et al.\ 2014;
Izmodenov \& Alexashov 2015).  Assuming these values for $V$ and $T$,
we can use equation~(1) to compute the value of $D$ necessary
to recover the $W=37.9^{\circ}$ value that is crudely
representative of the IBEX measurements.  The resulting divergence
is $D=0.94$.  We therefore would expect that the IBEX data could
also be reasonably well fit by a flow with parameters
($D$,$T$,$V$)=(0.94,21000,9), with a temperature that
is more plausible for the outer heliosheath than the $T=9500$~K
measurement of Kubiak et al.\ (2016).

     Obviously, an actual fit to the IBEX data needs to be performed
to verify that such a flow can reproduce the
observations.  Nevertheless, this exercise illustrates how the He
secondaries may provide a way to actually detect and measure the
divergence of the ISM flow around the heliopause.  This point becomes
even more valid if it can be demonstrated that a $D>0$ flow
demonstrably fits the IBEX data better than a $D=0$ flow.

     In future modeling of IBEX data using
a parametrized divergent flow, the model will
have to be fully 3-D, and it will be necessary to address the question
of how to deal with asymmetries in the flow caused by the ISM magnetic
field (e.g., Kubiak et al.\ 2014; Izmodenov \& Alexashov 2015).  One
approach is to simply allow the flow axis direction to differ from
the ISM flow direction, as Kubiak et al.\ (2016) have done.  The
apparent deflection of the secondary He flow from the primary flow
in a manner consistent with the ISM field orientation is in fact
one of the biggest lines of argument in favor of the ``Warm Breeze''
neutrals originating in the outer heliosheath (Kubiak et al.\ 2016).

     However, it is worth noting that the term ``deflection'' is not
the most physically accurate description of what is going on, neither
for the secondary He neutrals, nor for the secondary H neutrals
studied by Lallement et al.\ (2005).  There are, after all, no
neutrals that are actually being physically deflected at the
heliopause.  Instead, what is happening is an asymmetry in the
divergence of the flow, which yields an average velocity vector for
the flow inside the heliopause that is shifted from the original ISM
flow direction.  With a parametrized divergent flow it should actually
be possible to keep the central flow axis fixed to the flow direction
of the primary ISM flow component, but to allow the divergence to vary
with direction in some smooth fashion, relative to a plane defined by
the ISM magnetic field and flow directions.  Such an approach could in
principle provide a more physically realistic description of the flow
without increasing the number of free parameters of the fit.

     Regardless of how one implements a parametrized divergent
flow model, it is worthwhile to keep in mind that it will still only
be an approximation for the actual complex flow pattern.  It will
always be useful to use sophisticated global heliospheric
models (e.g, M\"{u}ller et al.\ 2013; Izmodenov \& Alexashov 2015)
to provide direct predictions for the He secondary flow properties
observed by IBEX.  Such models provide the most realistic
descriptions of the flow pattern in the outer heliosheath.  However,
a parametrized flow description is necessary to perform
model-independent fits to IBEX data, and in such fitting the
methodology proposed here should be preferable to the ``laminar
flow from infinity'' approximation.


\ack

Support for this project was provided by NASA award NNH16AC40I to
the Naval Research Laboratory.

\section*{References}
\begin{thereferences}
\item Bzowski, M., Kubiak, M. A, H{\l}ond, M., et al. 2014, A\&A, 569, A8
\item Bzowski, M., Kubiak, M. A, M\"{o}bius, E., et al. 2012, ApJS, 198, 12
\item Izmodenov, V. V., \& Alexashov, D. B. 2015, ApJS, 220, 32
\item Kubiak, M. A., Bzowski, M., Sok\'{o}l, J. M., et al. 2014, ApJS, 213, 29
\item Kubiak, M. A., Swaczyna, P., Bzowski, M., et al. 2016, ApJS, 223, 25
\item Lallement, R., Qu\'{e}merais, E., Bertaux, J. L., et al. 2005, Science,
  307, 1447
\item McComas, D. J., Allegrini, G., Bochsler, P., et al. 2009,
  Space~Sci.~Rev., 146, 11
\item McComas, D. J., Bzowski, M., Frisch, P., et al. 2015, ApJ, 801, 28
\item M\"{u}ller, H. -R., Bzowski, M., M\"{o}bius, E., \& Zank, G. P.
  2013, in Solar Wind 13, ed. G. P. Zank, et al. (New York:
  AIP, Vol.\ 1539), 348
\item Witte, M. 2004, A\&A, 426, 835
\item Wood, B. E., Karovska, M., \& Raymond, J. C. 2002, ApJ, 575, 1057
\item Wood, B. E., M\"{u}ller, H. -R., \& Witte, M. 2015, ApJ, 801, 62
\item Zank, G. P., Heerikhuisen, J., Wood, B. E., et al. 2013, ApJ, 763, 20
\end{thereferences}

\end{document}